# Markov-chain Monte-Carlo Sampling for Optimal Fidelity Determination in Dynamic Decision-Making

Abstract ID: 612372

Sara Masoud[1], Bijoy Chowdhury[1], Young-Jun Son[1], Russell Tronstad[2]

[1]Department of Systems and Industrial Engineering, University of Arizona

[2]Department of Agriculture and Resource Economics, University of Arizona

Tucson, Arizona, USA

## Abstract

Decision-making for dynamic systems is challenging due to the scale and dynamicity of such systems, and it is comprised of decisions at strategic, tactical, and operational levels. One of the most important aspects of decision-making is incorporating real-time information that reflects immediate status of the system. This type of decision-making, which may apply to any dynamic system, needs to comply with the system's current capabilities and calls for a dynamic data driven planning framework. Performance of dynamic data driven planning frameworks relies on the decision-making process which in return is relevant to the quality of the available data. This means that the planning framework should be able to set the level of decision-making based on the current status of the system, which is learned through the continuous readings of sensory data. In this work, a Markov-chain Monte-Carlo (MCMC) sampling method is proposed to determine the optimal fidelity of decision-making in a dynamic data driven framework. To evaluate the performance of the proposed method, an experiment is conducted, where the impact of workers performance on the production capacity and the fidelity level of decision-making are studied.

## 1. Introduction

Decision support tools that enable economical, effective, and real-time control of large-scale systems are central to their efficient operations. However, the complex and dynamic nature of such systems renders their coherent planning and control arduous. Decision-making for dynamic systems is challenging due to the scale and dynamicity of such systems, and it is comprised of decisions at strategic, tactical, and operational levels. One of the most important aspects of decision-making is incorporating real-time information that reflects immediate status of the system, which may create special problems [1]. Although entailing abilities such as dynamically incorporating data into an executing application simulation, steering measurement processes via simulation sound promising, there are some challenges in implementing such approaches [2]. First, the decisions have to be made in relation to demands of the environment. This real-time decision making needs to comply with the system's current capabilities and calls for a dynamic data driven planning framework. Second, both the system and the decision-making process seek to relate as integrated processes. As a result, multiple sensors can be employed within a dynamic system to report various quantities of interests within the system and contribute to real-time decision making. A third consequence is the need to consider the various fidelity levels of the dynamic decision-making processes. Performance of dynamic data driven planning frameworks relies on the decision-making process which in return is relevant to the quality of data available. This means that the planning framework should be able to set the level of decision-making based on the status of the system, which is learned through the continuous readings of sensory data.

A well-known instance of dynamic systems are production facilities, where continuous improvement of resource efficiency is a mandatory requirement for their survivors within the competitive global market. In any production facility, especially those concerning bioproducts, an optimized resource allocation can lead to a reduction in production costs, which in return sharpens the competitive edge of the product. In addition, layout design is directly related to resource allocation and line balancing. In labor-intensive bioproduction facilities such as grafting nurseries, resource allocation and layout design are especially complicated due to the uncertainties surrounding the dynamicity of workers' performance and bioproducts (i.e., young seedling plants).

In this work, a Markov-chain Monte-Carlo (MCMC) based Bayesian regression framework is proposed to obtain an optimal fidelity of decision-making in a dynamic data driven framework. From the Bayesian viewpoint, linear regression is modelled utilizing probability distributions rather than point estimates, and the response variable (i.e., y) is assumed to be drawn from a probability distribution. As a result, instead of finding the optimum value of model parameters, Bayesian regression determines the posterior distribution for the model parameters from raw data. What makes Bayesian methods so attractive is that it is fairly straightforward to adapt the model to challenging circumstances. Here, out of all MCMC algorithms, Gibb's sampling method is utilized to estimate the posterior distribution of the mentioned parameters. To evaluate the performance of the proposed method, multiple experiments were conducted, where the proposed method is embedded in our previously developed dynamic data driven adaptive simulation-based optimization (DDDASO) framework for a vegetable seedling propagation facility. The dynamic data driven framework is designed to handle decisions such as layout design, labor management, and irrigation scheduling for a vegetable seedling propagation nursery.

## 2. Methodology

Figure 1 displays the DDDASO framework proposed for real-time planning and control of the facility under study. The framework consists of three main units which are called the real system, measuring unit, and planning unit. The real system consists of the material, workers, management, computation units, and the sensors which are implemented to observe the behaviour of the system under study.

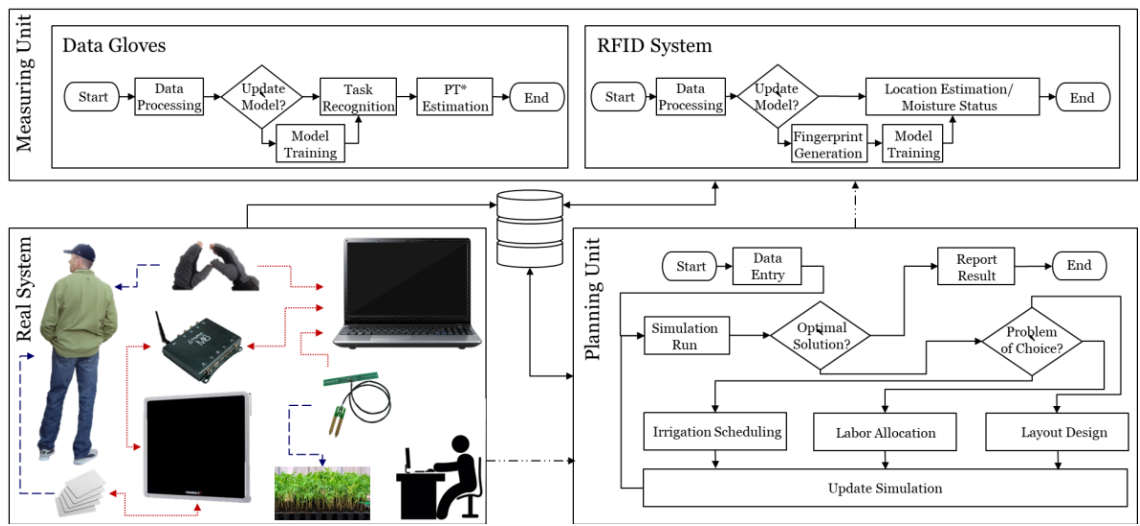

Figure 1: Dynamic data driven adaptive simulation-based optimization framework

The measuring unit handles the streams of sensory data and the estimation of processing time and material handling time of the system under study, details of which are discussed in [3] and [4], respectively. In addition, the measuring unit is responsible for detecting any anomaly in the system. The parameters estimated via the measuring unit are stored in a dataset and will be fed to the planning unit. The planning unit relies on the concept of simulation-based optimization where simulation mimics the performance of the system under study through the estimated parameters provided by the measuring unit, and optimization models look through different scenarios to find the optimal layout design [5], labor management [6], and irrigation scheduling. Although each one of problems has been discussed in great detail in different publications, the fidelity level of the dynamic decision making has not yet been addressed.

Layout design, labor management, and irrigation scheduling can be categorized as management decisions at strategic, tactical, and operational levels, respectively. As for the operational level problems (i.e., irrigation scheduling), the DDDASO framework calls for optimization in planning units as soon as an anomaly is detected via the measuring unit. As for layout design and labor management, it takes more than the detection of an anomaly to look for strategic or tactical changes. In other words, the system cannot go through optimizing the layout design or labor management every time a worker spends a longer time than expected to finish an assigned task. To address this issue, a Bayesian regression-based approach (i.e., Algorithm 4 in Figure 6) is developed where the connection between the short coming of production goals and the fidelity level of decisions are studied.

Bayesian regression facilitates uncertainties surrounding dynamic systems and represents such uncertainties within the model in contrast to most decision analyses based on maximum likelihood such as linear regression [7]. The Bayesian approach enforces the framework to look at historical data sets while updating the model based on incoming data. Figure 2 depicts a key idea of the Bayesian regression modeling.

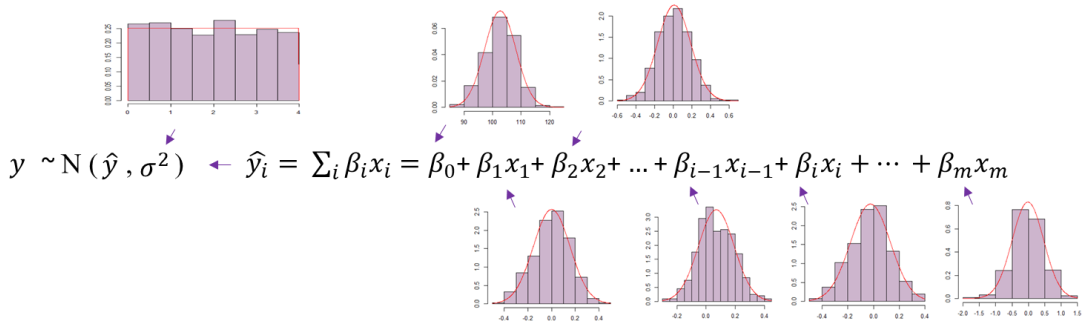

Figure 2: A key idea of Bayesian regression modeling

As shown in Figure 2, the linear relationship between the response variable y and predictive variables $X_1, X_2, ..., X_m$ is modelled by probability distributions rather than point estimates. As a result, y is generated from a normal (Gaussian) Distribution characterized by a mean and variance represented as $\sum_i \beta_i x_i$ and $\sigma^2$, respectively. The prior distribution of the coefficients shown in Figure 2, were defined by implementing Algorithm 1 which pseudocode is displayed in Figure 3.

---
Algorithm 1:

step 0: Read the historical dataset
for ( int i=1 in 1:1000):
    step 1: set the seed
    step 2: bootstrap a sample of size 500
    step 3: fit a linear model where y ~ b$_0$+b$_1$X$_1$+b$_2$X$_2$+ … +b$_i$X$_i$+… +b$_m$X$_m$
    step 4: update the coefficient vector B$_0$, B$_1$, …, B$_m$
    step 5: increment i
End for loop
step 6: fit distributions to B$_0$, B$_1$, …, B$_m$ for b$_0$, b$_1$, …, b$_m$
step 7: fit a distribution to variance of y

---

Figure 3: Algorithm 1- Estimating the prior distributions of the parameters of interest

Algorithm 1 defines the prior distributions of the coefficients by fitting one thousand linear regression models on bootstrapped datasets of size five hundred, where a linear regression model is fitted for each bootstrapped sample and the fitted parameters are stored in a separate matrix. After providing the one thousand estimated values for B$_0$, B$_1$, …, B$_m$, the algorithm fits distributions for each coefficient and the response variable as shown in Figure 1.

Given the prior distributions, the posterior distributions can be calculated by applying the Bayes rule to parameters of interest and the sensory data. The Bayes' rule states that the posterior distribution of the parameters of interest (e.g., B) given the incoming sensory data (i.e., D), can be written as $p(B|D) = p(D|B)p(B)/p(D)$. The posterior distribution (i.e., $p(B|D)$) provides the most complete information that is mathematically possible regarding the parameter values with respect the incoming sensory data. The only problem in calculating the posterior distribution is dealing with $p(D)$, which is also known as marginal likelihood and is defined as $\int p(D|B)p(B)dB$. Although for most cases closed mathematical forms cannot be obtained for the marginal likelihood, but MCMC sampling methods can be employed to accurately approximate the mentioned integration and provide the posterior distributions.

MCMC methods encompass a general framework of methods introduced by [8] for Monte Carlo integration, where $\int g(\theta)d\theta$ is estimated with a sample mean in which the original integration problem is defined as an expectation with respect to some density function *f(\*)* and MCMC generates samples from *f(\*)* by constructing a Markov Chain with stationary distribution *f(\*)*, and running the chain long enough to find the convergence of the chain to its stationary distribution. One of the MCMC sampling methods is Gibbs sampler which is a special case of Metropolis-Hasting sampler. Gibbs sampler usually outperforms other MCMC methods when the target is a multivariate distribution [9].

---
Algorithm 2:

step 0: initialize the chain at $\beta(0)$
Step 1: set the seed
for ( int i=1 in 1:5000):
    step 2: set $b_1 = \beta_1(t-1)$
    for ( int j=1 in 1:m+1): (for all the parameters of interest $b_0, b_1, \ldots, b_m, \sigma^2$)
        step 3: generates $\beta_j^*$ (t) from $f(\beta_j|x(-j))$
        step 4: update $b_j = \beta_j^*$ (t)
        step 5: increment j
    end for loop
step 5: set $\beta(t)= (\beta_1^*$ (t), $\beta_2^*$ (t), …, $\beta_{m+1}^*$ (t)
step 6: increment i
end for loop
step 7: burn the first 1000 chain and report the posterior distribution of y based on $b_0, b_1, \ldots, b_m, \sigma^2$

---

Figure 4: Algorithm 2- Estimating the posterior distribution of the production capacity (i.e., y)

To find the posterior of the parameters of interest (i.e., $b_0, b_1, \ldots, b_8, \sigma^2$), Algorithm 2 initializes the chain at state 0, $\beta(0)$ by giving the parameters some initializing values. Then, at each iteration of the chain, the algorithm draws a sample from the marginal distribution for each parameter and updates the chain based on the drawn samples. Given the impact of the initializing value for the parameters of interest, the algorithm discards the first one thousand observations and reports the results. Given the posterior distributions of the parameters, Algorithm 3, as shown in Figure 5, calculates the possibility of not meeting the targeted production goal if the no action is taken place.

---
Algorithm 3:

for ( int i=1 in 1:m):
    step 1: set $\bar{x}_i$
    step 2: increment *i*
End for loop
for ( int t=0 in 0:500 ):
    step 3: estimate $\bar{x}_i = \bar{x}_i + 0.1 * t * \bar{x}_i$, and update y distribution accordingly
    step 4: calculate P(Failure | $\overline{X(t)}$=( $\overline{x_1}, \overline{x_2}, \ldots, \overline{x_{j-1}}, \overline{x_j}, \ldots, \overline{x_{m-1}}, \overline{x_m}$))
    step 5: increment t
End for loop
step 6: define the relationship between P(Failure) and increase in X(t)

---

Figure 5: Algorithm 3- Impact of growth in material handling and processing time on production capacity

To calculate the possibility of not reaching the targeted production goal, the system captures the growth of processing and material handling times over duration of a shift by fitting the historical data and feeding the predicted values (i.e., X(t)) to the posterior distributions in order to calculate the possibility of not meeting the targeted production goal given a critical value. This critical value is decided through Algorithm 4 as shown in Figure 6.

---
Algorithm 4:

step 0: Read $m_{max}$ and D
step 1: Given the daily demand, optimize the labor allocation
step 2: Run Algorithm 1 to obtain prior distributions
step 3: Run Algorithm 2 to obtain the posterior distribution of production capacity (i.e., y)
step 4: Calculate $K_{critical} = (|(m_{max}*7*60/D)-D|/D)*100\%$
step 5: Run Algorithm 3 to find the P(Failure) for the critical value $K_{critical}$

Figure 6: Algorithm 4- Defining the critical value

Algorithm 4 brings together the DDDASO framework and Algorithms 1, 2, and 3. As in the first step, the algorithm reads the maximum capacity of hiring grafting workers (i.e., $m_{max}$) given the optimal layout design and daily demand (i.e., D). In the next step, Algorithm 4 runs the simulation-based labor allocation module, where the optimal labor allocation plan is decided for the daily demand, and simulated data are provided to feed Algorithm 1. Next, Algorithm 1 runs to provide prior distributions for the parameters of interest. Utilizing prior distributions, Algorithm 2 will define the posterior distribution of the production capacity (i.e., y). Finally, Algorithm 3 is utilized to find the probability of failure for the critical value (i.e., $K_{critical} = (|(m_{max}*7*60/D)-D|/D)*100\%$).

## 3. Experiment and Results

The experimental setup in this work is defined based on a seedling propagation facility, where a daily production goal of 100 trays is targeted. In addition, given the layout design of the facility a total of 30 workers can be hired to graft the 100 trays. Here, $\hat{y}_l = \sum_i \beta_i x_i = \beta_0 + \beta_1 x_1 + \beta_2 x_2 + \beta_3 x_3 + \beta_4 x_4 + \beta_5 x_5 + \beta_6 x_6 + \beta_7 x_7 + \beta_8 x_8$ and the $x_i$s, as the value of the observed parameters, are the time estimations of the main grafting processes (i.e., scion cutting, rootstock cutting, rootstock clipping and joining) and the major material handling activities (i.e., healing to growing, growing to grafting, grafting to healing, and healing to growing) in a grafting operation. By feeding the observed $x_i$s to the Algorithm 1, the following prior distributions have been fitted for the parameters of interest $b_0, b_1, \ldots, b_8, \sigma^2$ as *N(108, 2.12²), N(-0.0004, 0.15²), N(-0.010, 0.17²), N(-0.07, 0.12²), N(-0.0125, 0.11²), N(-0.025, 0.15²), N(-0.127, 0.25²), N(-0.015, 0.48²), N(-0.006, 0.15²), u(0.001, 3.99)*, respectively. Figure 5 displays the Q-Q plots for the fitted distributions to visualize the goodness of fit.

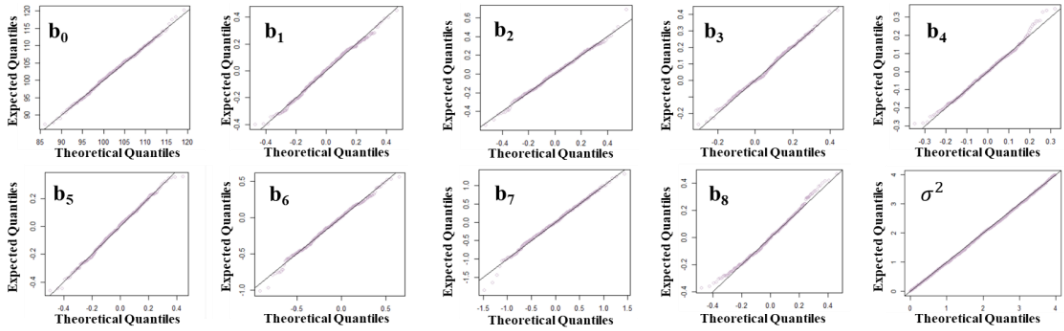

Figure 7: Q-Q plots, visualizing the goodness of fit for the fitted prior distributions

By implementing Algorithm 2 and utilizing the Bayes' rule the posterior distribution of y is defined as *N($\hat{y}$, 3.34)*, where $\hat{y}$ is defined as following.

$$\hat{y} = 104 - 0.0222X_1 - 0.0221X_2 - 0.0164X_3 - 0.0229X_4 - 0.0046X_5 - 0.00131X_6 - 0.0053X_7 - 0.0086X_8$$

By implementing the maximum capacity (i.e., $m_{max}$) of 30 and a daily demand (i.e., D) of 100 trays within the formula (i.e., $K_{critical} = (|(m_{max}*7*60/D)-D|/D)*100\%$), a critical value (i.e., $K_{critical}$) of 26% is achieved. In the next step, the distribution of the *N ($\hat{y}$, 3.34)* and the estimated average processing time and martial handling times are given to Algorithm 3. Algorithm 3 defines the possible growth in material handling and processing times as $\overline{x}_l = \overline{x}_l + 0.1 * t * \overline{x}_l$, $\forall i \in \{1,2,\ldots,8\}$ and $t \in [0,100]$ and calculates the probability of failure given the estimated increase in values of the mentioned variables (i.e., $P(Failure| \overline{X(t)}=(\overline{x_1}, \overline{x_2}, \overline{x_3}, \overline{x_4}, \overline{x_5}, \overline{x_6}, \overline{x_7}, \overline{x_8}))$). Figure 8 displays the impact of growth in processing and material handling times in relation to the probability of a failure.

Given the relationship between the probability of failure and growth in processing and material handling times, 26% of the increase in material handling and processing times can lead to failure with a probability of 0.625 if no necessary action is taken, as the maximum increases by reoptimizing the labour allocation. If the increase in material handling and processing times exceeds the critical value of 26%, the labor management optimizer will not be effective anymore and a new optimization of layout design and resource allocation is required. For the given range of [27%, 100%] the P(Failure) is provided via Figure 8.

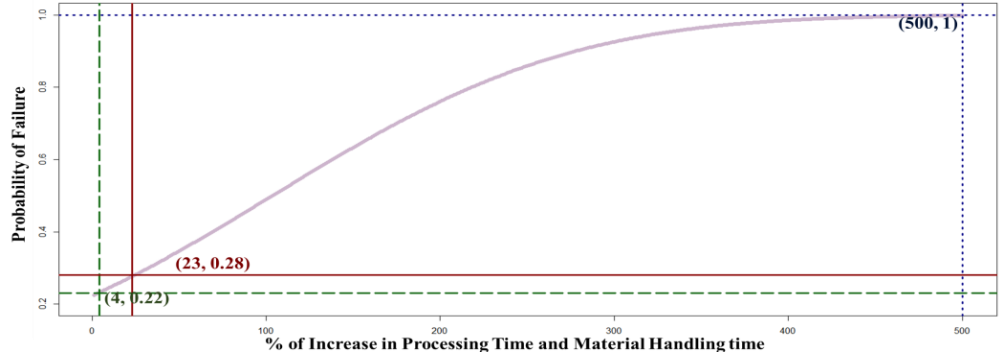

Figure 8: Mapped relationship between the % increase in material handling and processing time and P(Failure)

As a result, the framework will call for optimizing the labor management if any increase between 1% and 26% for the average processing time of the facility under study is observed. Such increase in probability of failure is a result of an average reduction of 2 to 25 trays in production capacity due to the possibility of an unbalanced production line and bottlenecks within the system as a result of increased average processing and material handling times and can be handled by reoptimizing the labor management problem. For any increase of 26% or more, the system needs to reoptimize the layout and resource optimization problem.

## 4. Conclusions

Dynamic decision making, defined as a series of independent actions that must be taken over time to monitor and control a system's status, can improve the performance of the system in terms of productivity and efficiency if the decision making takes place at an appropriate time. To achieve that, four algorithms (i.e., Algorithms 1,2, 3 and 4) have been developed to set the fidelity level of decision making in a dynamic data driven planning and control environment. Through these algorithms, a relationship among the average increase in processing and material handling time and the probability of not meeting the designated production goal is defined, where the performance of the workers is monitored by the DDDASO framework and the proposed algorithms predict the need of reoptimizing labor management or layout design problems to prevent shortages in the presence of anomalies.

## Acknowledgements
This work has been supported by U.S. Department of Agriculture (USDA) – National institute of food and agriculture under project number 2016-51181-25404.